\documentclass[preprint]{aastex631}

\usepackage{amsmath}
\newcommand{\bma}[1]{\boldsymbol{#1}}
\newcommand{\unit}[1]{\ensuremath{\,\mathrm{#1}}}

\shorttitle{AASTeX v6.3.1 Sample article}
\shortauthors{Zhou et al.}
\graphicspath{{./}{figures/}}

\begin{document}

\title{Frozen-field Modeling of Coronal Condensations with MPI-AMRVAC \\ I: Demonstration in two-dimensional models}

\correspondingauthor{Yuhao Zhou}
\email{yuhao.zhou@kuleuven.be}

\author[0000-0002-4391-393X]{Yuhao Zhou}
\affiliation{Centre for mathematical plasma-astrophysics (CmPA), KU Leuven \\
Celestijnenlaan 200B, 3001 Leuven, Belgium}

\author[0000-0001-8164-5633]{Xiaohong Li}
\affiliation{Max Planck Institute for Solar System Research \\ 
Göttingen D-37077, Germany}
\affiliation{Centre for mathematical plasma-astrophysics (CmPA), KU Leuven \\
Celestijnenlaan 200B, 3001 Leuven, Belgium}

\author[0000-0003-3544-2733]{Rony Keppens}
\affiliation{Centre for mathematical plasma-astrophysics (CmPA), KU Leuven \\
Celestijnenlaan 200B, 3001 Leuven, Belgium}



\begin{abstract}
Large-scale coronal plasma evolutions can be adequately described by magnetohydrodynamics (MHD) equations. 
However, full multi-dimensional MHD simulations require substantial computational resources. 
Given the low plasma $\beta$ in the solar corona, in many coronal studies, it suffices to approximate the magnetic field to remain topologically fixed and effectively conduct one-dimensional (1D) hydrodynamic (HD) simulations instead.
This approach is often employed in studies of coronal loops and their liability to form condensations related to thermal instability.
While 1D HD simulations along given and fixed field line shapes are convenient and fast, they are difficult to directly compare with multi-dimensional phenomena. Therefore, it is more convenient to solve volume-filling, multi-dimensional versions of the MHD equations where we freeze the magnetic field, transforming it into frozen-field HD (ffHD) equations for simulation. 
We have incorporated this ffHD module into our open-source MPI-AMRVAC code and tested it using a two-dimensional (2D) evaporation--condensation model to study prominence formation due to radiative losses. 
The 2D ffHD results are compared with those from actual 2D MHD and pseudo-2D HD simulations, analyzing the differences and their causes.
Pseudo-2D studies account for the known  {flux tube} expansion effects.
Overall, the performance of 2D ffHD is close to that of 2D MHD and pseudo-2D HD. 
The 2D tests conducted in this paper will be extended in follow-up studies to 3D simulations based on analytical or observational approaches.
\end{abstract}



\section{Introduction} \label{sec1}
The solar corona is largely composed of ionized plasma, thus we typically employ single-fluid magnetohydrodynamic (MHD) equations to describe large-scale activities in the solar corona.
For example, the ideal MHD momentum equation, which describes motion as influenced by inertia, pressure gradients, Lorentz forces and gravity, is expressed as:
\begin{equation}
\rho \frac{d\bma{v}}{dt} =-\nabla p+\bma{J}\times \bma{B}+\rho \bma{g},
\label{eq11}
\end{equation}
where all the symbols have their usual meaning.
Due to the large magnetic Reynolds numbers in the corona, it is generally considered that the motion of coronal plasma follows the ideal MHD frozen-in paradigm \citep{alfv1943}.
This means that the magnetic field and electrically conducting fluids move together.

Furthermore, the solar corona is an environment characterized by high temperature and low density, with its plasma $\beta$ (as the ratio of plasma to magnetic pressure) often being less than one.
As a result, coronal activities are predominantly governed by the magnetic field, and hence we often can focus on the magnetic field evolution alone, forgetting about the plasma component (by e.g. setting $\beta=0$). 
Then, actual resistive MHD effects can induce magnetic reconnection which alters the topology of magnetic fields, and both ideal and resistive MHD effects may trigger instabilities and subsequent eruptive events \citep[see, for example][]{amar1996, anti1999a, chen2011, shib2011, liu2020, li2021}.

On the other hand, in both quiet Sun regions and in active regions, there are also long-term (compared with the time scale of a thermodynamic phenomenon under study) stable structures, such as solar filaments or coronal loops. 
These persistent stable structures are often considered to be close to a force-free state.
This assumes that in a magnetically dominated scenario, only structures where the Lorentz force is near zero can stably exist over extended periods.
In such cases, every magnetic flux tube can be regarded as a rigid body.
{In such cases, the magnetic field force is infinitely strong, so there are no motions of the plasma due to this infinite magnetic force, and therefore no effect in energy and momentum equations. As a result, every magnetic flux tube can be regarded as a rigid body.}
This in effect reduces the problem to 1D HD simulations to investigate field-guided evolutions along stable magnetic field structures over long durations or to perform simulations of short-lived phenomena. 

A typical example where this approach proved useful is the study of filament formation due to evaporation--condensation.
This is a well-known means to explain filament/prominence formation, and this is hypothesized to occur in magnetic flux ropes of varying size which connect opposite polarity magnetic footpoint regions.
By heating the footpoints of such a flux rope at heights including the upper chromosphere to transition region, plasma is evaporated into the solar corona, triggering thermal instabilities and leading to condensation \citep{park1953, anti1999b}. 
To model the bare essentials of such runaway radiatively-driven thermodynamics, the motion of the magnetic field is neither taken into consideration nor required. 
Using our open-source MPI-AMRVAC\footnote{https://www.amrvac.org/}~\citep{xia2018, kepp2023}, a number of such 1D HD studies have been performed \citep{xia2011, zhou2014, huan2021}. 
Similar works have also been done by other codes \citep[for example, ][]{from2018, pelo2022} in recent years.

Another application for 1D fixed-field simulations involves the study of solar flares using RADYN \citep{carl2023} or FLARIX \citep{vara2010} codes.
In such flare-relevant 1D HD evolutions, one can take full account of advanced non-local thermal equilibrium radiative effects, which would otherwise be inhibiting in full multi-D MHD settings.
These various variants of fixed-field HD simulations have been successful to a certain extent in explaining physical phenomena and replicating observational results.
Indeed, if the magnetic topology can be considered unaltered for the phenomenon of interest, purely 1D simulations capture the crux of the problem and can successfully explain certain physical phenomena.

The greatest advantage of 1D simulations lies in their computational speed and efficiency. 
Of course, 1D results often cannot be effectively represented in multi-dimensional space – you always have just a 1D magnetic flux tube.
Turning 1D studies into actual synthetic observables along given lines-of-sight will always require assumptions on the flux tube cross-section and its embedding environment.
An intermediate step to multi-D evolutions can consider the so-called pseudo-multi-dimensional simulation, which involves extracting hundreds of magnetic field lines from a given multi-dimensional topology and then interpolating the results of the one-dimensional simulation back into the multi-dimensional space. 
This simple and effective method has achieved considerable results, as exemplified by \citet{luna2012} and \citet{guo2022}.
However, while the actual 1D simulation time is short, the initial extraction of magnetic field lines, e.g. obtained from a magnetofrictional evolution to an actual force-free 3D magnetic field) and the subsequent interpolation process can be time-consuming, sometimes even comparable to the duration of the simulation itself, and can be affected by interpolation errors. 
Moreover, this pseudo-multi-D approach faces issues related to the filling factor – how many magnetic field lines to extract and how to appropriately interpolate them back into the original space are matters of discussion. 

A more realistic multi-dimensional treatment is the so-called frozen-field method first adopted by \citet{mok2005}, which involves conducting fluid dynamic simulations in a multi-dimensional setting, assuming magnetic flux freezing and a rigid magnetic topology, purely restricting the direction of fluid flow. 
Compared to complete MHD simulations, this method significantly reduces the computational resources required and does not have some of the data processing issues present in pseudo-multi-dimensional simulations, making it simpler and more convenient, consistent with standard multi-dimensional simulations. 
Since the magnetic field is frozen in place, we refer to this method as frozen-field HD (ffHD).

In \citet{mok2005, mok2008}, this method is primarily used to study the large-scale thermal evolution of active regions and coronal loop structures as affected by thermal instability, with extreme ultraviolet (EUV) images synthesized through forward modeling compared with observations.
Indeed, this method actually solves the governing equations on a volume-filling grid, identical to regular multi-dimensional MHD simulations, and hence can be readily turned into synthetic observables.
\cite{mok2016} updated this model by using a heating mechanism inspired from Alfv{\'e}n wave turbulence dissipation.
\citet{lion2013} also used this model for similar studies, mainly analyzing the impact of the coronal heating model on the results. 
\citet{wine2014} utilized this model to diagnose the physical parameters of coronal loops, and investigated the time lags between different EUV channels \citep{wine2016}.
However, there has been no study on large-scale filament structures and their formation due to evaporation-condensation using this approach.

Therefore, this paper will introduce and demonstrate the novel ffHD physics module of our open-source MPI-AMRVAC toolkit, first proving its effectiveness through a variety of 2D tests.
Note that our implementation inherits the full dimension-independent Adaptive Mesh Refinement capability of the MPI-AMRVAC toolkit.
Subsequently, in later articles, we will apply it to three-dimensional (3D) simulations of filament dynamics.
The structure of this paper is as follows: Section \ref{sec2} introduces the assumptions behind and implementation details of the ffHD equations. 
In Section \ref{sec3}, we conduct two-dimensional (2D) tests under solar atmospheric conditions, followed by a detailed comparison between 2D ffHD model, pseudo-2D  {model, and true 2D MHD runs in Section \ref{sec4}.}
Finally, in Section \ref{sec5}, we provide a discussion of the different models and summarize our conclusions.

\section{Equations}
\label{sec2}
\subsection{MHD equations}\label{sec21}
We start from the conservative ideal MHD equations including for the moment only external gravity as a source term:
\begin{eqnarray}
    \frac{{\partial \rho }}{{\partial t}} + \nabla  \cdot (\rho \bma{v}) & =  & 0,
    \label{eq21}\\
    \frac{{\partial (\rho \bma{v})}}{{\partial t}} + \nabla  \cdot (\rho \bma{vv} + {p_{tot}}\bma{I} - {\bma{BB}}) & = & \rho \bma{g},
    \label{eq22}\\
    \frac{{\partial e}}{{\partial t}} + \nabla  \cdot \left( {e\bma{v} + {p_{tot}}\bma{v} - \bma{BB} \cdot \bma{v}} \right) & = & \rho \bma{g} \cdot \bma{v},
    \label{eq23}\\
    \frac{{\partial {\bma{B}}}}{{\partial t}} + \nabla  \cdot (\bma{vB - Bv}) & = & \mathbf{0}.
    \label{eq24}
\end{eqnarray}
We use density $\rho$, external gravitational acceleration $\bma{g}$, velocity vector $\bma{v}$, and magnetic field vector $\bma{B}$.
Here, $e=p/(\gamma-1)+\rho \bma{v}^2/2+\bma{B}^2/2$ is total energy density, while $p_{tot}=p+\bma{B}^2/2$ is total pressure.
A fully ionized atmosphere, where the number density ratio between Hydrogen and Helium ($n_H : n_{He}$) is 10:1, is assumed in this work.
 {Based on this assumption}, we have $\rho = 1.4 n_H m_p$ and the ideal gas law takes the form $p = 2.3 n_H k_B T$, where $m_p$ is the mass of a proton and $k_B$ is the Boltzmann constant, which relates pressure, density and temperature $T$.

\subsection{Field-Frozen HD equations}\label{sec22}
As in \citet{mok2005, mok2008} and following works, the equations of the ffHD method is not explicitly listed, and the equations in those works were written in the primitive form, we will, for the first time, discuss the derivation of the conservative form of the ffHD equation here.

If we assume the magnetic field is static, while the field line acts as a guiding rigid structure, we have,  {in addition to the MHD equations}, the following relations:
\begin{eqnarray}
    \frac{\partial \bma{B}}{\partial t} & = & \mathbf{0}\,, 
    \label{eq25}\\
     \frac{\partial \bma{\hat{b}}}{\partial t} & = & \mathbf{0}\,, 
    \label{eq25hat}\\
    \bma{v} & = & v_\parallel \bma{\hat{b}},
    \label{eq26}
\end{eqnarray}
where $\bma{\hat{b}}$ is the unit vector along the fixed magnetic field.
 {Notice should be taken that, handling magnetic null points, where $B=0$ could be challenging.
Nevertheless, given the assumption that the magnetic field lines are static, we have the flexibility to slightly adjust the grid positions to circumvent these null points.}

By using Eq.~(\ref{eq25}) and Eq.~(\ref{eq26}), we can simply drop the entire induction equation Eq.~(\ref{eq24}), as it is obeyed trivially.
And by combining Eq.~(\ref{eq21}) and Eq.~(\ref{eq25}), the continuity equation writes as
\begin{equation}
    \frac{\partial \rho}{\partial t}+\nabla \cdot  \left (  \rho v_{\parallel} \bma{\hat{b}}\right ) = 0.
\label{eq27}
\end{equation}

When we write the momentum equation Eq.~(\ref{eq22}) using the current density vector $\bma{J}=\nabla \times \bma{B}$, we obtain
\begin{equation}
    \frac{\partial \left(\rho \bma{v}\right)}{\partial t} + \nabla \cdot \left(\rho \bma{v} \bma{v}\right) + \nabla p - \bma{J} \times \bma{B} =\rho \bma{g}.
    \label{eq29}
\end{equation}
Projecting Eq.~(\ref{eq29}) onto $\bma{\hat{b}}$ and using Eq.~(\ref{eq26}) will lead to
\begin{equation}
    \bma{\hat{b}} \cdot \frac{\partial \left(\rho v_{\parallel} \bma{\hat{b}}\right)}{\partial t} + \bma{\hat{b}} \cdot \left[ \nabla \cdot \left(\rho v_{\parallel} v_{\parallel} \bma{\hat{b}} \bma{\hat{b}}\right)\right] + \bma{\hat{b}} \cdot \nabla p = \rho \bma{\hat{b}} \cdot \bma{g}\,.
    \label{eq210}
\end{equation}
Note that the Lorentz force is orthogonal to $\bma{\hat{b}}$ and hence it no longer appears in this equation.



 {For any scalar S, we can get $\nabla \cdot \left(S\bma{\hat{b}}\bma{\hat{b}}\right) = S\bma{\hat{b}} \cdot \nabla  \bma{\hat{b}} + \bma{\hat{b}} \nabla \cdot \left(S\bma{\hat{b}}\right)$. Since $\bma{\hat{b}}$ is a unit vector, we have $\bma{\hat{b}}\cdot \bma{\hat{b}}=1$. Therefore, the first term is zero because $\bma{\hat{b}} \cdot \nabla \bma{\hat{b}} = \nabla \left(\bma{\hat{b}} \cdot \bma{\hat{b}}\right)/2 = 0 $.
Multiple by $\bma{\hat{b}}$ on the left then we can get  $\bma{\hat{b}}\cdot \left[ \nabla \cdot \left(S\bma{\hat{b}}\bma{\hat{b}}\right)\right] = 
\bma{\hat{b}} \cdot \left(\bma{\hat{b}} \nabla \cdot \left(S\bma{\hat{b}}\right)\right)=\left(\bma{\hat{b}}\cdot \bma{\hat{b}}\right)\nabla\cdot\left(S\bma{\hat{b}}\right) =\nabla \cdot (S \bma{\hat{b}})$.}

We can use this to manipulate Eq.~(\ref{eq210}) directly into 

\begin{equation}
    \frac{\partial \left(\rho v_{\parallel}\right)}{\partial t} + \nabla \cdot \left[\rho v_{\parallel}^2 \bma{\hat{b}}\right] +  \bma{\hat{b}}\cdot \nabla p =\rho g_{\parallel},
    \label{eqcorrect}
\end{equation}
where $\rho g_{\parallel}$ is the field-aligned component of the gravity force, i.e. $g_{\parallel}=\bma{\hat{b}}\cdot\bma{g}$. 

We then find that we can write it as
\begin{equation}
    \frac{\partial \left(\rho v_{\parallel}\right)}{\partial t} + \nabla \cdot \left[(\rho v_{\parallel}^2 +p) \bma{\hat{b}}\right] -p (\nabla\cdot \bma{\hat{b}})=\rho g_{\parallel}.
    \label{eqcorrectB}
\end{equation}

For the energy equation, we first define the purely hydrodynamic total energy density $E=p/\left(\gamma-1\right) + \rho v_{\parallel}^2/2$ such that $e=E+\bma{B}^2/2$.
Starting from Eq.~(\ref{eq23}), we can have,
\begin{equation}
    \frac{\partial E}{\partial t} + \nabla \cdot \left[\left(E+{\bma{B}^2/2}\right)v_{\parallel}\bma{\hat{b}} + \left(p+{\bma{B}^2}/2\right)v_{\parallel}\bma{\hat{b}} -{|\bma{B}|^2\bma{\hat{b}}\bma{\hat{b}}} \cdot v_{\parallel}\bma{\hat{b}}\right] = \rho g_{\parallel} v_{\parallel}.
    \label{eq214}
\end{equation}
Then, we can have the new energy equation,
\begin{equation}
    \frac{\partial E}{\partial t} + \nabla \cdot \left[\left(E + p\right) v_{\parallel} \bma{\hat{b}}\right] = \rho g_{\parallel} v_{\parallel}.
    \label{eq215}
\end{equation}
In this way, we have the ffHD equations,
\begin{eqnarray}
    \frac{\partial \rho}{\partial t}+\nabla \cdot  \left (  \rho v_{\parallel} \bma{\hat{b}}\right ) &=& 0,
    \label{eq216}\\
    \frac{\partial \left(\rho v_{\parallel}\right)}{\partial t}+\nabla \cdot \left [  \left (  \rho v^2_{\parallel} +p\right ){\bma{\hat{b}}}\right ] &=& \rho g_{\parallel} + p (\nabla\cdot \bma{\hat{b}}),
    \label{eq217}\\
    \frac{\partial E}{\partial t} + \nabla \cdot \left[\left(E + p\right) v_{\parallel} \bma{\hat{b}}\right] &=& \rho g_{\parallel} v_{\parallel}.
    \label{eq218}
\end{eqnarray}

Thus far, the energy equation we discussed did not include non-adiabatic effects, such as thermal conduction or radiative losses.
When full non-adiabatic energy equation is considered, we will have:
\begin{equation}
    \frac{\partial E}{\partial t} + \nabla \cdot \left[\left(E + p\right) v_{\parallel} \bma{\hat{b}}\right] = \rho g_{\parallel} v_{\parallel} + \nabla  \cdot {\bma{q}}- RC + H.
    \label{eq219}
\end{equation}

The parallel thermal conduction term  {$\nabla \cdot \bma{q}$} is treated in the same way with gravity, and calculated by using the Super-time-stepping method \citep{meye2012}, where $\bma{q} = \kappa (\bma{\hat{b}} \cdot \nabla T) \bma{\hat{b}}$ and $\kappa=10^{-6}T^{5/2}$~erg cm$^{-1}$ s$^{-1}$ K$^{-1}$ is the Spizer-type heat conductivity.
The optically thin radiative cooling term $RC= 1.2{n_\mathrm{H}}^2\Lambda \left( T \right)$ is calculated using the exact integration scheme from \citet{town2009}, based on the tabulated radiative cooling table  {\citep{colg2008, xia2011}}.
The heating term $H$ will be introduced later.

To capture the evaporation and energy exchange in a more accurate way, transition region adapitve conduction \citep[TRAC,][]{john2019, zhou2021} method is also implemented in this module.
However, in the present work, it is not activated for a better comparison.
 {For the same reason, adaptive mesh refinement (AMR) is not activated.}

\subsection{pseudo-multi-D equations}
\label{sec23}
From the derivation above, we know that this ffHD model is supposed to be in-between the pseudo-multi-D HD model (i.e. a collection of 1D simulations along selected and fixed field lines) and the MHD model.
For the pseudo-multi-D model,  {due to its 1D nature}, we solve the following 1D HD equations \citep[see][for derivation]{shod2018}:
\begin{eqnarray}
\frac{{\partial \rho}}{{\partial t}} + \frac{1}{A} \frac{\partial }{{\partial s}}(A\rho v) &= & 0\,,
\label{eq38}\\
\frac{\partial (\rho v)}{{\partial t}} + \frac{1}{A} \frac{\partial }{{\partial s}}\left(A(\rho {v^2} + p)\right) &= & \frac{p}{A} \frac{dA}{ds} + \rho {g_\parallel}\,,
\label{eq39}\\
\frac{{\partial E}}{{\partial t}} + \frac{1}{A} \frac{\partial }{{\partial s}}\left(A\left(E + p\right)v \right) &=& \rho {g_\parallel}v + \frac{1}{A} \frac{\partial }{{\partial s}}\left(A\kappa \frac{{\partial T}}{{\partial s}}\right) -RC + H\,,
\label{eq310}
\end{eqnarray}
where $s$ represents the 1D coordinate along the field line.
The factor $A(s)$ is the expansion factor that describes the change of the cross section of the 1D flux tube along the field line.

In earlier 1D simulations,  {both} models incorporating either a constant $A$ \citep{anti1999b, xia2011}  {and} a variable $A(s)$ \citep{karp2005, miki2013, from2018} have been successfully employed to simulate condensation phenomena in magnetic flux bundles of given and fixed shape.
Here we adopt both assumptions.
 {For convenience, we label the pseudo-multi-D simulations with constant $A$ as HD and those with variable A as HDexp.}

 {In Table~\ref{tb1}, we list the models mentioned above, which will be used in the present work for quick comparison.}

\begin{longtable}{c p{7cm} cc}
\caption{Different models used in the present work}\label{tb1}\\
\hline
Label & Description & Variables & Equations \\
\hline
\endfirsthead
\hline
\endfoot
 MHD & full MHD simulations& $\rho,\bma{v},p,\bma{B}$ & Eq.~(\ref{eq21})--(\ref{eq24}) + source terms\\
 ffHD & multi-D HD simulations, but the motion is field line aligned & $\rho,v_{\parallel},p$  & Eq.~(\ref{eq216})--(\ref{eq217}), Eq.~(\ref{eq219})\\
 HD & pseudo-multi-D HD simulations, aggregation of 1D HD simulations with constant $A$ & $\rho, v, p$ & Eq.~(\ref{eq38})--(\ref{eq310})\\
 HDexp & same as HD, but with variable $A$ & $\rho, v, p$ & Eq.~(\ref{eq38})--(\ref{eq310})\\
\end{longtable}

\section{Application in 2D simulations}
\label{sec3}
\subsection{Solar atmosphere setup}
\label{sec31}
Now, we  {first} apply the ffHD equations in a typical 2D solar atmospheric simulation.
The basic setup is similar to the one adopted in our previous works \citep{zhou2021,zhou2023}.
The 2D simulation domain ranges -50 Mm$<x<$50 Mm and 0$<y<$80 Mm with a resolution of 960$\times$768 uniform grids.
Each cell has a size of 104 km $\times$ 104 km.
The initial temperature distribution is based on the classic VAL-C model \citep{vern1981} and the coronal part is extrapolated based on the assumption that the heat flux is constant.
 {Then, by setting a bottom (at $y=0$) Hydrogen number density of $7.1 \times 10^{14}$ cm$^{-3}$, we can have the initial density and pressure distribution by assuming hydrostatic equilibrium.
The initial velocity is set to zero.} 
The magnetic field is an analytic potential field described by the following equations:
\begin{eqnarray}
    {B_x}  & =  & + \pi {B_{\mathrm{0}}}\cos \left( {\frac{{\pi x}}{{2{L_0}}}} \right){e^{ - \frac{{\pi y}}{{2{L_0}}}}} - \pi {B_{\mathrm{0}}}\cos \left( {\frac{{3\pi x}}{{2{L_0}}}} \right){e^{ - \frac{{3\pi y}}{{2{L_0}}}}}\,,
    \label{eq31} \\
    {B_y} & = &  - \pi {B_{\mathrm{0}}}\sin \left( {\frac{{\pi x}}{{2{L_0}}}} \right){e^{ - \frac{{\pi y}}{{2{L_0}}}}} + \pi {B_{\mathrm{0}}}\sin \left( {\frac{{3\pi x}}{{2{L_0}}}} \right){e^{ - \frac{{3\pi y}}{{2{L_0}}}}}\,.
    \label{eq32}
\end{eqnarray}
Here, $L_0$ is 50 Mm.
In the ffHD model, the magnetic field is static and only its local direction is involved in the dynamic evolution.
Thus, the choice of $B_0$ which sets the magnitude of the coronal field has no influence on the ffHD results.
This is obviously different for full 2D MHD simulations, where the vector $\mathbf{B}$ can evolve, and where flow can have a component locally perpendicular to the magnetic field lines (displacing them according to the frozen-in theorem).

Our space- and time-dependent heating source $H(x,y,t)$ as adopted in the energy equation Eq.~(\ref{eq219}) is a weak exponential background heating, in addition to a stronger but localized heating term, described as 
\begin{eqnarray}
    H\left(x,y,t\right) &=& H_{\mathrm{bgr}}\left(y\right)+H_{\mathrm{loc}}\left(x,y,t\right)\,,
    \label{eq33} \\
    H_{\mathrm{bgr}}\left(y\right) &=& H_0\exp{\left(y/\lambda_0\right)}\,,
    \label{eq34} \\
    {H_{\mathrm{loc}}\left(x, y, t\right)} &=& {H_1}{R_{\mathrm{{ramp}}}}\left(t\right)H_x\left(x\right)H_y\left(y\right)\,,
    \label{eq35} \\
    H_x\left(x\right) &=& \exp \left( { - \frac{{{{\left( {x - {x_r}} \right)}^2}}}{{{\sigma ^2}}}} \right) + \exp \left( { - \frac{{{{\left( {x - {x_l}} \right)}^2}}}{{{\sigma ^2}}}} \right)\,,
    \label{eq36} \\
    H_y\left(y\right)  &=& \exp \left( { - \frac{{{{\left( {y - {y_1}} \right)}^2}}}{{\lambda _1^2}}} \right).
    \label{eq37}
\end{eqnarray}
Parameters are chosen the same as in our previous work \citep{zhou2023}: $H_0=3 \times 10^{-4}~\unit{erg}\unit{cm^{-3}}\unit{s^{-1}}$, $H_1=2 \times 10^{-2}~\unit{erg}\unit{cm^{-3}}\unit{s^{-1}}$, $\lambda _0 = 50$~Mm, $x_l= -41.5$~Mm, $x_r= 41.5$~Mm, $\sigma = 6.32$~Mm, $y_1 = 4$~Mm, and $\lambda _1 = 6.32$~Mm.
$R_\mathrm{{ramp}}$ is a ramp function making $H_\mathrm{loc}$ increase gradually, changing from 0 to 1 linearly in 10 minutes.
As for boundary conditions, we simply set all boundaries closed.

Though the initial condition is force balanced, since it combines a potential magnetic field with a 1D hydrostatic atmosphere, this state does not represent a thermal dynamic equilibrium state.
Indeed, the anisotropic thermal conduction will redistribute the heat along the field topology, while the exponential background heating versus the temperature- and density-dependent radiative cooling will need to settle the configuration to a thermal and force balance. This is true for all simulations, whether done in full MHD, or in an pure ffHD setup where the magnetic field can not respond.
Therefore, before we start the simulation (before $t=0$), we relax the system for 450 minutes, which is also true to all the other simulations mentioned in the present work.
Figure \ref{fig1}(a1) shows the temperature distribution right after this relaxation for the ffHD run, i.e. at $t=0$.
We also overplot in panel (a1) some field lines (black and white solid lines) to represent the configuration of the magnetic field.

After the relaxation, we start to introduce the localized heating term $H_\mathrm{loc}$.
Figure \ref{fig1}(a2) shows the obtained temperature distribution at $t=143$ min for the ffHD model.
We can clearly see the formation of a central prominence-like condensation, demonstrating already a basic validation of the ffHD method.
In what follows, we will discuss the essential differences with other approaches (the other panels in Figure ~\ref{fig1}) in more details.

\subsection{Comparison of the ffHD approach with other models}\label{sec32}
\begin{figure*}
  \centering
  \includegraphics[width=0.95\textwidth]{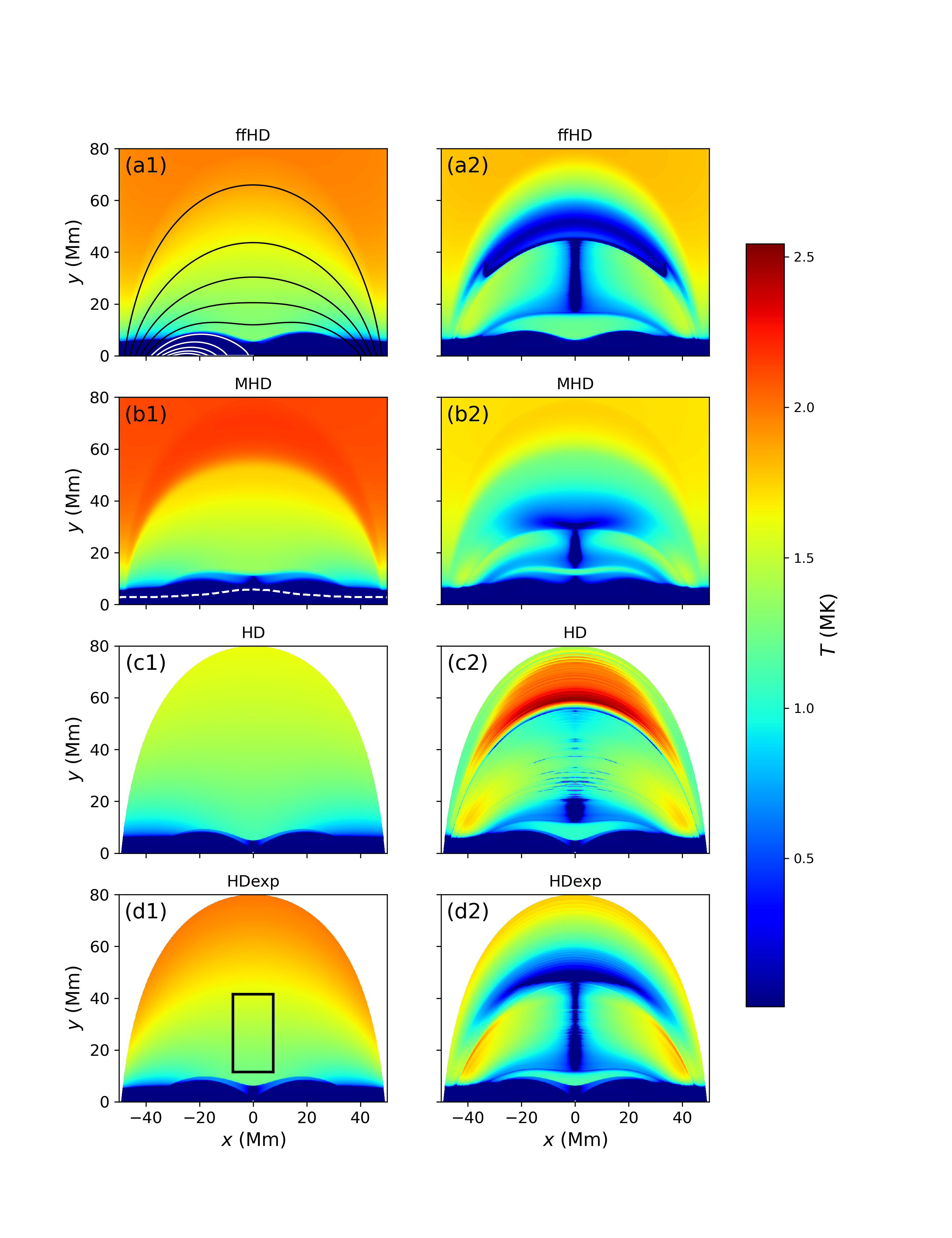}
  \caption{(a1--d1) Temperature distribution at $t=0$ (the end of our relaxation phase) for ffHD, MHD, HD and HDexp models, respectively.
  (a2--d2) The same at $t=143$ min after added localized footpoint heating.
  The solid lines in panel (a1) shows the configuration of magnetic field.
  The dashed line in panel (b1) shows the position where plasma $\beta=1$.
  The rectangle in panel (d1) shows the select region that will be used in Figure~\ref{fig2}.}
  \label{fig1}
\end{figure*}

 {Following the ffHD simulation, we perform MHD and pseudo-2D simulations to examine the differences.}
The first one is a full 2D MHD simulation, which enables the deformation of magnetic field lines.
 {The MHD simulation solves Eq.~(\ref{eq21})--(\ref{eq24}) with the extra source terms as in the ffHD equations.}
 {The initial setup of the MHD simulation, }followed by a relaxation phase and then a phase with localized background heating, in this 2D MHD simulation is done analogously as for the ffHD case. 
$B_0$ is chosen to be 2 G here.
This choice will be discussed in the following section, since unlike any of the ffHD or pseudo-2D HD approaches, the MHD outcome can be influenced by the initial field strength.
A background field splitting method \citep{xia2018} is applied for the MHD case.
Figure~\ref{fig1}(b1) and (b2) shows the temperature distribution of the MHD simulation at $t=0$ and $t=$143 min, respectively.
We find the condensation is smaller and at a lower position, which comes from the fact that the development of the condensation is slower compared with the ffHD model.

For the related pseudo-2D HD simulations, we first need to extract several field lines from the 2D domain.
These field lines are expected to cover the simulation domain as completely as possible.
Considering that the resolution of the 2D simulation is 960$\times$768, we chose 960 seed points on the $y=0$ axis.
To avoid double count at the left-right symmetry, these seed points are not selected equidistantly at grid centers or grid edges, but a little bit deviated.
And then, we integrate 960 field lines from these seed points.
However, since the field lines integrated from seed points near the two lower corners will go outside the top boundary of the 2D simulation domain, we decided not to use these incomplete field lines.
As a result, only 953 field lines are taken for our pseudo-2D simulations, and they all represent closed arcade field lines as visualized in Figure~\ref{fig1}(a1), which can be either dipped or undipped centrally.

The initial atmosphere and heating functions are interpolated from the 2D setup onto the given set of field lines, making the initial conditions of the various simulations as similar as possible.
The boundary conditions for the collection of 1D simulations are kept the same with the bottom boundary condition of the 2D simulation.
We adapt the number of the uniform grids of each 1D simulation so that the resolution will be more or less 100 km.
However, for short field lines, such a resolution choice will result into too few cells, which makes the simulation unstable.
Therefore, for field lines whose length is lower than 3.2 Mm, we always use 32 grids to conduct the simulation.
After all the 1D simulations are done (the 953 runs are completely independent of each other, so can be done in embarrassingly parallel fashion), we interpolate all the field lines back onto the 2D domain.

 {As mentioned in Sect.~\ref{sec23}, in previous works, both constant and variable $A$ are adopted.
Here, we first show the result with the constant $A=1$ pseudo-2D model, which is labeled as case HD.
In this case, the actual cross-sectional variation of the fixed field is not accounted for at all.}
In Figure~\ref{fig1}(c1) and (c2), we show the temperature distribution at $t=0$ and $t=143$ min, respectively.
Note the missing region in top left and right corners due to our deliberate selection of only field lines that close within our domain.

In the rightmost two panels of Figure~\ref{fig1}, the corresponding results of a varying $A$ are shown.
$A$ is here chosen to be proportional to $1/|\bma{B}|$ (inverse of the local field strength), as in many previous works.
{As mentioned,} we label this case as HDexp.
By comparing Figure \ref{fig1}(a2) and Figure \ref{fig1}(c2), we can see that the HDexp model exhibits quite similar results to the volume-filling ffHD model.
However, the HD model in Figure~\ref{fig1}(d2) shows a clearly different scenario.
This is consistent with the findings of previous 1D studies, pointing to the importance of the field line geometry \citep{miki2013}.

\begin{figure*}
  \centering
  \includegraphics[width=0.95\linewidth]{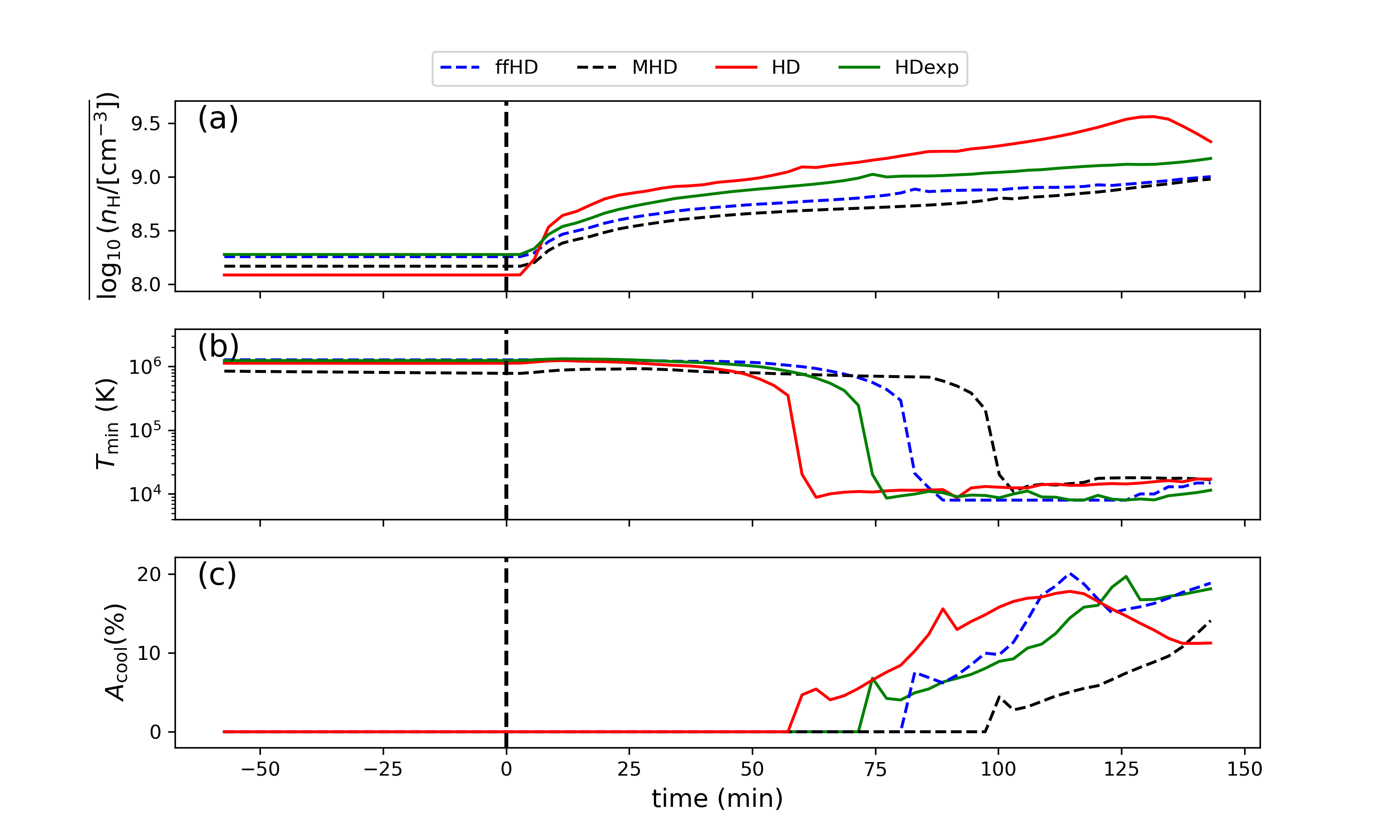}
  \caption{Time evolution of (a) averaged number density  {(calculated with the mean of $\log_{10}(n_\mathrm{H})$ values)}, (b) minimum temperature $T_{\mathrm{min}}$ and (c) area percentage of the condensations $A_\mathrm{{cool}}$ of the selected region in Figure \ref{fig1}(d1), for all four models. The vertical dashed line indicates the starting of localized heating, e.g., $t=0$.}.
  \label{fig2}
\end{figure*}

To analyze the similarity more quantitatively, we choose a rectangular region as marked in Figure~\ref{fig1}(d1).
We trace the time evolution of the averaged number density $\overline {n_\mathrm{H}}$ within this region, and plot the result for all four simulations in Figure~\ref{fig2}(a).
 {Note that, because the number density within the region spans over two orders of magnitude, we did not directly take the average of the number density; instead, we took the average of $\log_{10}(n_\mathrm{H})$.}
The evolution of this centrally averaged number density can show the speed of evaporation.
The vertical dashed line indicates the turning on of our localized heating, i.e. our $t=0$.
As expected, the ffHD model (in blue dashed line) is always in-between the HDexp (in green solid line) and the full MHD (in black dashed line) model, whereas the MHD model is overall evolving slower, with less density accumulated centrally due to the footpoint evaporations.
However, the HD model (in red) shows a much quicker mass accumulation.
Note that the pre-additional heating state is slightly different for each case, but has the typical coronal density values.

In Figure~\ref{fig2}(b), we show the time evolution of $T_{\mathrm{min}}$, the minimum temperature reached within this same rectangular region, for these four models.
$T_\mathrm{min}$ can indicate the onset of thermal instability{, typically associated with the sudden decline in its value to values near 10000 K}.
The actual minimum reachable is influenced by the cooling curve adopted, and by the lower temperature tretament of the cooling curve, as investigated in idealized local box setting by \citet{herm2021}.
Again, the ffHD model and HDexp model show a very similar evolution.
The condensation for the MHD is the slowest while the HD model is evolving the fastest.

Figure~\ref{fig2}(c) illustrates the percentage of the total rectangular area occupied by cold matter, characterized as having a temperature $T<0.05$~MK.
Again, we see that the HD model has the fastest condensation, and the MHD model is slower than the others.
The ffHD model has a similar evolution with the HDexp case.

According to the above analysis, we qualitatively conclude that the results of HDexp, ffHD and MHD model are similar, all exhibiting condensation and the formation of filaments in a similar way. 
The condensation in the HD model occurs the fastest, while in the MHD model, it is the slowest.
The results of ffHD and HDexp are comparatively close.
This is reasonable, as the fundamental assumptions underlying both simulations are similar, with discrepancies primarily from algorithmic and discretization and interpolation errors.
Regarding the MHD model, due to the dynamical involvement of magnetic fields, the results are slightly different. 
 {The precise cause of such difference remains unclear, but it may be related to the fact that these HD models account only for acoustic waves, whereas the MHD model encompasses a variety of wave modes with distinct velocities.}
However, they are broadly consistent and do form a filament in a combination of dipped and undipped arcade field lines, due to chromospheric evaporation occuring at the footpoints.
In the following, we will do a detailed comparison and investigation of the differences between these four cases, as well as explore the reasons behind these variations.

\section{Further Analysis}
\label{sec4}
\subsection{Pseudo-2D runs: HD versus HDexp}
\label{sec41}
From the results of the simulations presented above, it is evident that there is a difference between the HDexp model and HD model in the evaporation--condensation simulations.
In magnetic field configurations where the magnetic flux changes rapidly, or in other words, where the cross-sectional area of the magnetic flux tube varies significantly, an even greater discrepancy might be observed. 

Physically, HDexp is expected to be more consistent. 
This is because, to ensure the conservation of magnetic flux, the cross-sectional area of the higher-altitude magnetic flux tubes naturally increases as the magnetic field strength decreases. 
 {However, as mentioned above, in many previous 1D simulations, the constant-area assumption is chosen.
This is not only because constant-area is computationally or numerically simple, but also comes from the fact that the coronal magnetic fluxes appear constant in width or cross section area in the actual coronal image data from observations.}
Therefore, both models have been successful in generating evaporation--condensation phenomena. 
This raises a question: How significant is the impact of varying cross-sectional areas on the simulation of evaporation-condensation processes?

\citet{miki2013} investigated this question in detail.
They found that an HDexp model is more likely to experience thermal non-equilibrium (TNE) behaviour than the HD model.
The coupling between the chromosphere, transition region, and the coronal loop part indeed can show so-called TNE behaviour, as the overall timescales involved in all (non-)adiabatic processes may compete to never achieve an actual steady (thermally and force-balanced) endstate \citep{klim2019}.
We will look into this question from a different aspect.
We first select a 1D magnetic field line, specifically the one with its footpoints at $x=\pm 45$~Mm.
This magnetic field line has a length of 124 Mm, with its footpoints near the localized added heating center. 
This particular field line shows condensations in both HD and HDexp models.
The ratio between the maximum and minimum cross section  {of the flux tube} along this field line is about 9:1, which quantifies the effect of area variation as being one order of magnitude.
With the magnetic  {flux tube} shape fixed, the primary parameters influencing the localized heating are the intensity of the heating $H_1$, its location (altitude $y_1$) and the scale height $\lambda_1$.
The impact of the heating location ($y_1$) on condensation is related to the configuration and is not discussed here, as we keep it fixed at 4 Mm, a little below our relaxed height of the transition region. 
Instead, we focus on the intensity and scale height of the heating. 
Using this selected magnetic field line, we conducted a parametric survey by varying the $H_1$ and $\lambda_1$ to observe the differences, especially the differences in the condensation process, between HD and HDexp under various heating conditions.

We use the minimum temperature within the coronal region, $T_{st} = \min\left(T\left(s,t\right)\right)$, where $s_l<s<s_r$, and $0<t<t_c$, to quantify whether a condensation happened on the field line within the time $t<t_c$,
 {, where $t_c$ is the current endtime, }$s_l$ and $s_r$ are the left and right boundaries of the coronal region.
They are determined from the initial $t=0$ temperature variation only, according to the criterion $T(s_l<s<s_r, t=0)>1$~MK.
Note that this coronal length segment is the same for all runs, as the initial variation is not yet influenced by the added localized heating.

\begin{figure*}
  \centering
  \includegraphics[width=0.95\linewidth]{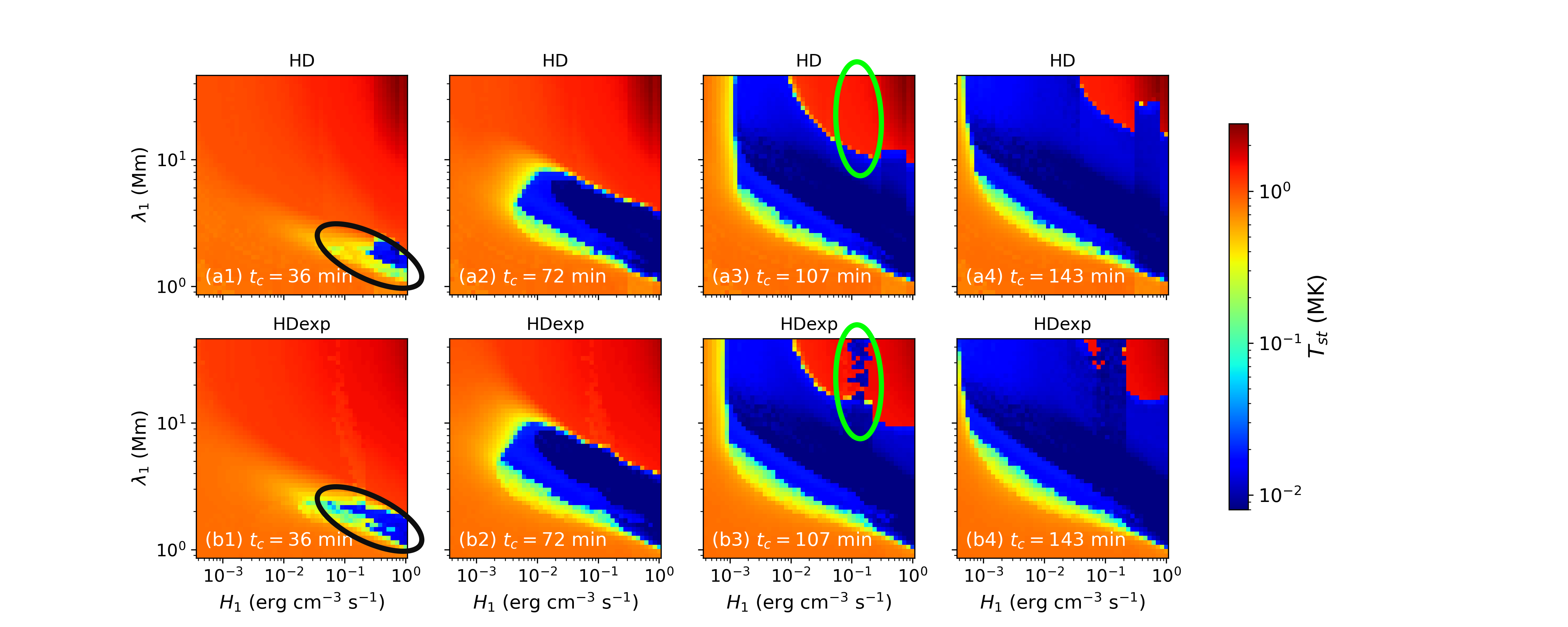}
  \caption{Distribution of the minimum coronal temperature $T_{st}$ with respect to localized heating parameters $H_1$ and $\lambda_1$ in the HD model at endtime $t_c=$(a1) 36 min, (a2) 72 min, (a3) 107 min and (a4) 143 min, respectively.
  The corresponding results for the HDexp model are shown in the lower 4 panels.  {Ellipses are regions where significant differences exist.}}
  \label{fig3}
\end{figure*}

 {Figure~\ref{fig3} is composed of 5000 individual 1D simulations, e.g., 2500 runs for the HD case and another 2500 runs for the HDexp cases.}
In panels (a1)--(a4), we display the distribution of $T_{st}$ for the HD model at four endtimes, namely $t_c=36$, 72, 107, and 143 minutes, respectively.
The horizontal and vertical axes represent the heating intensity $H_1$ and heating scale height $\lambda_1$, respectively.
Here, $H_1$ varies from 4$\times$10$^{-4}$ erg cm$^{-3}$ s$^{-1}$ to 1 erg cm$^{-3}$ s$^{-1}$, and $\lambda_1$ from 0.89 Mm to 44.7 Mm.
Panels (b1)--(b4) depict the corresponding scenarios for the HDexp model.
Note that the steps of $H_1$ and $\lambda_1$ is chosen to be logarithmic.

A comparison of panels (a1) and (b1) reveals that, under conditions of higher heating intensity and lower scale height  {(the area marked by the ellipse)}, the HDexp model tends to form condensations earlier than the HD model.
Under the same conditions, the parameter space for the HD model to form condensations is somewhat smaller, which is basically consistent with \citet{miki2013}.

From panels (a2) and (b2), it is evident that both HDexp and HD are more prone to condensation with increasing heating intensity. However, the `optimal' scale height (where condensation forms most rapidly) varies with different heating intensities. 
The higher the heating intensity, the shorter the `optimal' scale height.
This relationship appears approximately linear on our double logarithmic scale. 
At $t=72$ minutes, the range of conditions under which the  HDexp model forms condensations is still slightly larger than that of the HD model, but the difference is minimal. 
A similar pattern is observed in panels (a3) and (b3), except in the top-right corner  {(the area marked by the ellipse)}, near $H_1=0.15$ erg cm$^{-3}$ s$^{-1}$, where the  {overall heating is extremely strong}.
Still, the conditions for condensation formation are roughly similar for both models. 
By $t=143$ minutes, most of the parameter space has already experienced condensation (panels (a4) and (b4)), except for some regions in the lower left and upper right. 
The main difference is observed in the upper right, where under conditions of higher heating intensity and large scale height, the tendency to form a condensation varies.

\begin{figure*}
  \centering
  \includegraphics[width=0.95\linewidth]{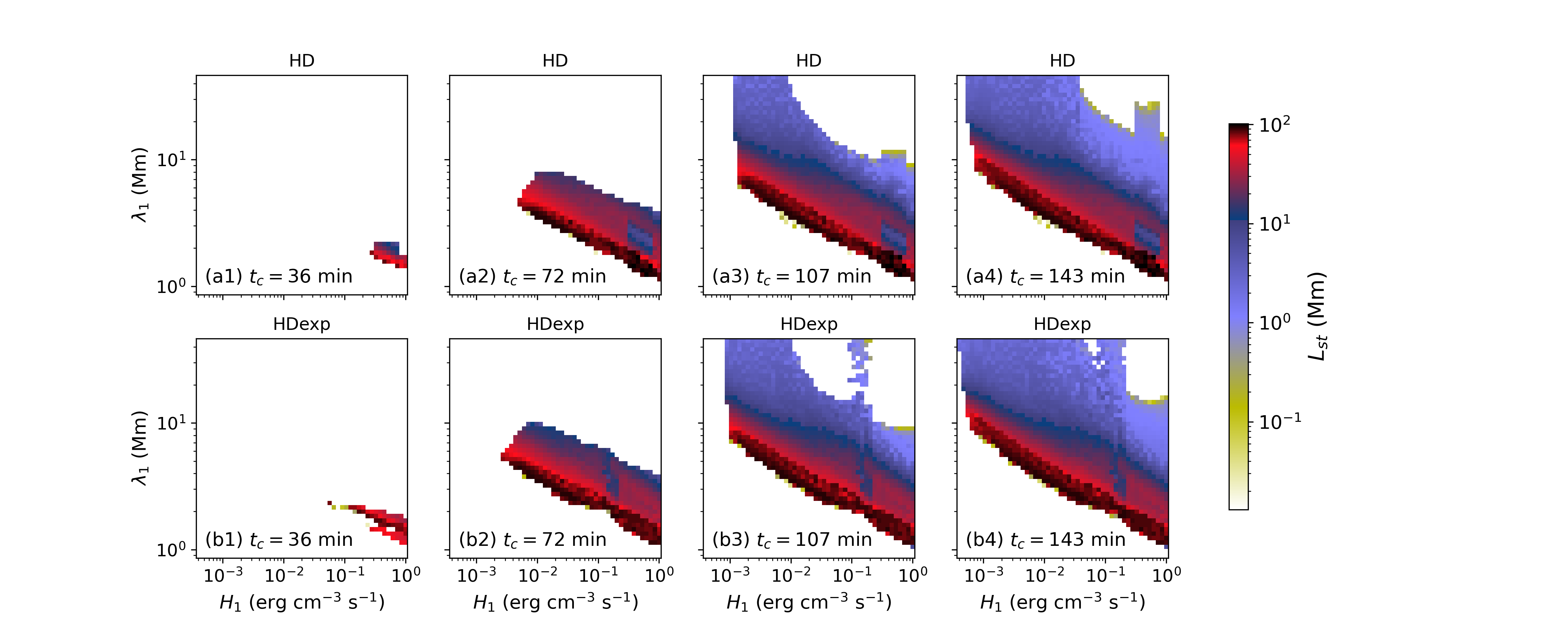}
  \caption{Same as Figure \ref{fig3}, but now showing the distribution of $L_{st}$, the longest length of the condensation with these parameters.}
  \label{fig4}
\end{figure*}

While $T_{st}$ indicates the occurrence of condensation, it does not describe the full thermodynamic state after condensation.
One aspect of this state is the actual length achieved by the centrally forming condensation.
We use $L_{st}$, the maximum length of the condensation (where $T<0.05$~MK) in the corona during the time interval $0<t<t_c$, to characterize it.
Figure \ref{fig4} shows the result. 
Similar to Figure \ref{fig3}, panels (a1)--(a4) depict the distribution of $L_{st}$ for the HD model at four endtimes $t_c=36$, 72, 107 and 143 minutes.
The remaining panels correspond to the HDexp model. 
It is evident that Figures \ref{fig3} and \ref{fig4} highly correlate.
Areas where condensation forms earlier typically have lower temperatures and longer filaments, and vice versa for areas where condensation forms later. 
However, the length of the filaments is similar, for both HD and HDexp models, across most of the parameter space.

Therefore, for the selected magnetic field line, there is no significant difference in physical outcomes between the HDexp and HD models for different heating parameters. 
This is somewhat in contrast with the findings by \cite{miki2013}.
However, significant differences were observed in our pseudo-2D simulations at higher locations, suggesting that some magnetic field lines are more sensitive to changes in heating parameters and more prone to exhibit differences. 
These field lines are often longer and exhibit more significant changes in cross-sectional area (with ratio between the maximum and minimum cross section goes up to over 20), potentially leading to more pronounced differences in the rate of condensation formation.

\subsection{Full 2D runs: ffHD versus MHD}
\label{sec42}
As mentioned earlier, and visually evident when comparing panels Figure~\ref{fig1}(a2) and (d2), the difference between the pseudo-2D HDexp model and the full 2D ffHD model is minor.
This difference between pseudo-2D and frozen field HD approaches has already been explored by \cite{mok2005}.
Thus, here we focus on the difference between another group, the frozen-field ffHD model and the MHD model, which are both simulations done in actual multi-dimensional settings.

As mentioned in Section~\ref{sec1}, the underlying assumption taken to use ffHD as a proxy for an MHD evolution is due to the low plasma $\beta$ values in the solar corona, where the plasma motion is unlikely to significantly affect the magnetic field. 
However, it is important to note that the bottom part of our simulation region includes the chromosphere and the transition region, where the density is relatively high, resulting in larger thermal pressure. 
In some areas of our simulation domain, the plasma $\beta$ even exceeds one.
In Figure~\ref{fig1}(b1), we overplot a dashed line to show the position where plasma $\beta=1$.
Consequently, particularly at the lower atmosphere associated with the evaporation process, this assumption that the field can not be influenced by the plasma motion is not valid. Therefore, it is a not surprise that we can find variations in the evaporation--condensation process between both 2D models.

Let us take a closer look at the lower regions, especially in areas where the evaporation process occurs, to determine the extent of the differences between ffHD and MHD models.
Firstly, it is important to recognize that differences between the ffHD and MHD models already emerge before introducing additional local heating, i.e. already such differences exist in the relaxed states at our $t=0$, as evident from comparing Figure~\ref{fig1}(a1) and (b1). 
Here, we select two sets of magnetic field lines for analysis at $t=0$. 
Observing the configuration of the magnetic field (e.g., Figure \ref{fig1}(a1)), it is evident that there are two types of magnetic loops in this system: the `long loops' spanning both sides of the simulation area (e.g., the black lines), and the `short loops' confined to either the left ($x<$0) side or the right ($x>$0) side (e.g., the white lines).
These loops exhibit distinctly different physical processes.

\begin{figure*}
  \centering
  \includegraphics[width=0.99\linewidth]{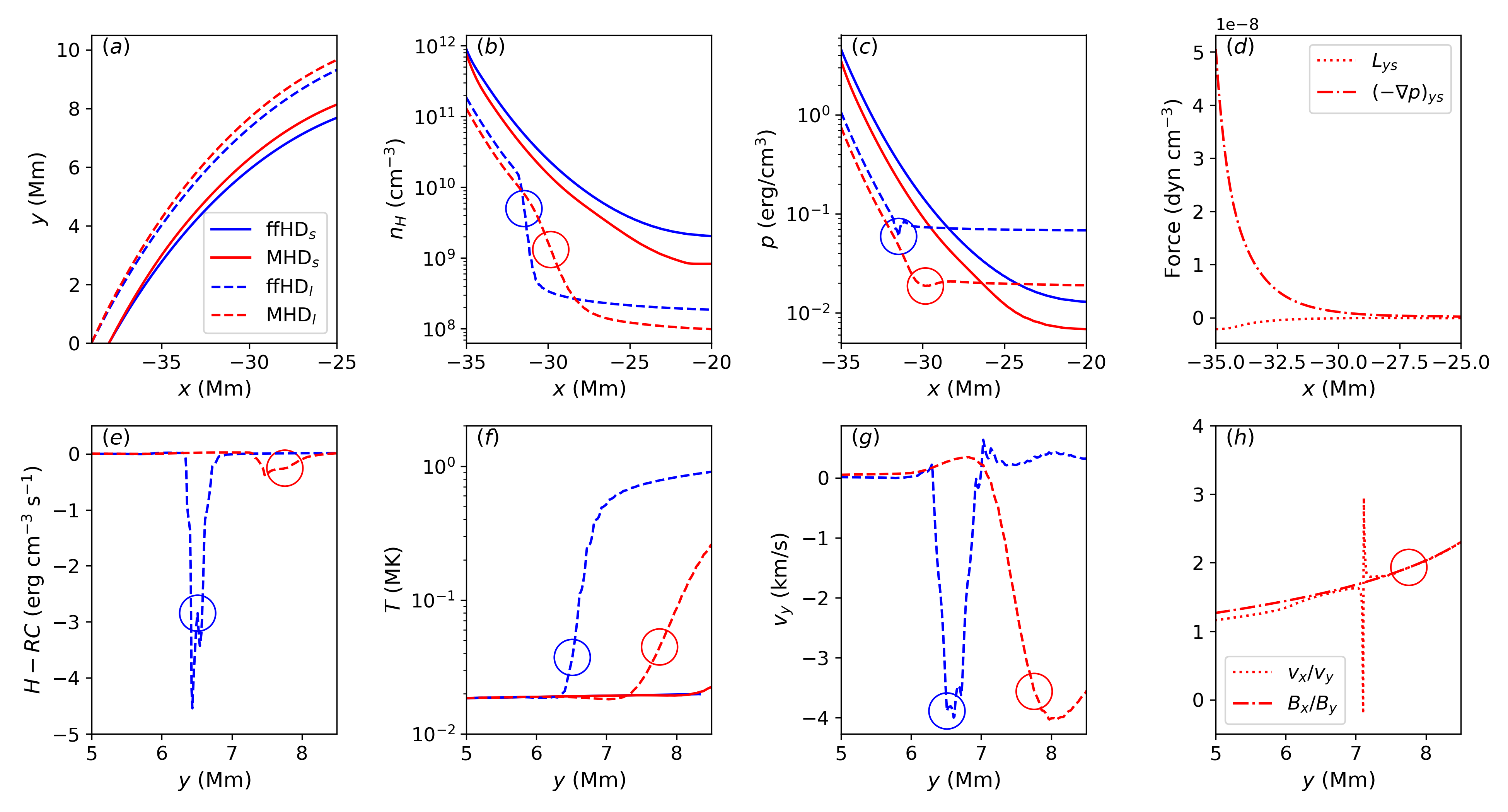}
  \caption{(a) The configuration of the magnetic field lines, (b) the number density distribution, (c) the gas pressure distribution, and (f) the temperature distribution of both short and long loop ffHD$_l$, ffHD$_s$, MHD$_l$, MHD$_s$ cases, respectively. (d) The $y-$ direction forces for the short loop MHD$_s$ case; (e) the net energy gain $H-RC$ and (g) velocity in the $y-$direction $v_y$ for the long loop MHD$_l$ and ffHD$_l$ cases. (h) The ratio of the velocity $v_x/v_y$ versus the ratio of magnetic field  {strength} $B_x/B_y$ for the long loop MHD$_l$ cases. The blue and red circles correspond to the locations of the minima in the long loop pressure distribution.}
  \label{fig5}
\end{figure*}

The separatrix line between the `short' and `long' loops has footpoints located approximately at $x=\pm 38.5$~Mm.
Therefore, we choose two sets of magnetic field lines with footpoints at $x=-39$~Mm and $x=-38$~Mm.
We present the part of the magnetic field lines near these footpoints in Figure \ref{fig5}(a).
The lines are colored in red for the MHD model and in blue for the ffHD model, with solid lines representing short loops and dashed lines representing long loops, and they will be denoted with subscripts $s$ and $l$ as in ffHD$_{s,l}$ and MHD$_{s,l}$, respectively. 
Despite the proximity of their footpoints, the length of the short loop is around 42 Mm, while that of the long loop is about 85 Mm, nearly double the length of the short one. 
In the ffHD model, the magnetic field lines are fixed, while in the MHD model, they undergo certain deformation after relaxation, slightly expanding outwards.
A similar effect, but on the localized heating, is investigated in our previous work \citep{zhou2023}.

Figure \ref{fig5}(b) shows the number density $n_H$ distribution along these two sets of magnetic field lines. 
Note the logarithmic scale on the vertical axis, indicating that in the corona, the density difference between the ffHD and MHD models can be significant. 
For instance, at $x=-32.5$~Mm, the number density in the ffHD$_s$ case is 1.10$\times$10$^{11}$ cm$^{-3}$, whereas in the MHD$_s$ model it is 7.31$\times$10$^{10}$ cm$^{-3}$, the former being 50\% more than the latter.
The higher the magnetic field lines, the more pronounced this disparity. 
At the apex of the field lines, the density in the ffHD$_s$ case is nearly three times that of the MHD$_s$ case.
This density difference is also true for the long loops, namely between the ffHD$_l$ and MHD$_l$ cases. 
Thus, after relaxation, the coronal densities in the two models already show significant differences. 
This inevitably leads to a slower condensation in the evaporation phase for the MHD model due to insufficient coronal density.
The corresponding pressure distribution is shown in Figure \ref{fig5}(c), where we observe a similar situation, with the pressure in the MHD model being several times lower than in the ffHD model.

Then, is the difference in density (and pressure) merely due to the slight expansion of the magnetic loops which occurs in MHD, while it is excluded in ffHD? 
In fact, for the short loops, the expansion of the magnetic  {loops} can indeed fully explain this density difference.
Considering the position at $x=-32.5$~Mm, the height corresponding to the ffHD$_s$ case is $y=4.53$~Mm, and the density of the MHD$_s$ case at $y=4.53$~Mm is 1.04$\times$10$^{11}$ cm$^{-3}$, which is already close to the 1.10$\times$10$^{11}$ cm$^{-3}$ of the ffHD$_s$ case.

The remaining differences can be attributed to the (vertical component of the) Lorentz force $L_y$, which by construction only plays a role in the full MHD case.
We know that in this system, the vertical force balance is approximately determined by $(-\nabla p)_y = \rho g + L_y$.
The initial magnetic field is a force-free potential field, so after the expansion we obtain due to the relaxation up to $t=0$, there must be an inward Lorentz force.
The existence of this force effectively weakens the pressure gradient force, resulting in a lower coronal pressure in the MHD model, and hence a lower density. 
Figure \ref{fig5}(d) shows the distribution of the Lorentz force in the MHD$_s$ case, confirming this point (dotted line represents the Lorentz force $L_y$, and dash-dotted line represents the pressure gradient force $(-\nabla p)_y$), where the Lorentz force indeed partially offsets the pressure gradient force for the short loops.

However, for the long loops, the issue is not so straightforward. 
While the two aforementioned factors (i.e. geometric difference due to expansion and the non-zero Lorentz contribution for MHD only) do play a role, other reasons also contribute. 
For the long loops, the coronal density of the MHD$_l$ case is much lower than that of the ffHD$_l$ case, but its pressure and density distributions are much more complex than those of the short loops.
Two very localized dips appear in the pressure distribution along the long loops, as seen in panel Figure~\ref{fig5}(c). 
On the one hand, at their position marked with circles, there is a region with a local pressure minimum; 
on the other hand, in the region $x < -25$~Mm where we have coronal conditions, the pressure of the higher-positioned long loops is actually greater than that of the lower-positioned short loops. 
The local pressure dip is related to our model, and mainly coming from the optically thin radiative loss treatment. 
In panel (e), we display $H - RC$, the net energy increase obtained by subtracting radiative losses from background heating (ignoring the effect of thermal conduction). 
It can be seen that at the position of the minimal pressure, there is a significant radiative loss. 
This is influenced by the sharp variation in the radiative loss curve at low temperatures which we already know is also affecting the actual thermodynamic evolutions \citep{herm2021}. 
However, we argue that this does not fully explain the lower coronal density and pressure of the MHD$_l$ case compared to the ffHD$_l$, as both the ffHD and MHD models exhibit this minimal pressure region.

A pronounced difference is also evident in the temperature distribution. 
Let us first examine the temperature distribution (Figure \ref{fig5}(f)), where we
note that we now use height $y$ (instead of $x$) as the horizontal axis.
It can be seen that after relaxation, the short loops have almost completely cooled to 1.8$\times$10$^4$~K (the cutoff temperature of our radiative loss curve), while the higher parts of the long loops can still reach around 1~MK.
As this temperature increase signals the transition region height, the transition region of the ffHD$_l$ case starts at around a height of $y = 6.1$~Mm, while that of the MHD$_l$ case extended to about 7.1 Mm.
It is this difference in the position of the transition region that leads to the several-fold difference in pressure seen in panel Fig~\ref{fig5}(c) and the density disparity in panel (b).

The question then becomes why is there a significant difference in the height of the transition region between the two models after a period of relaxation? 
We argue this is because, in the MHD$_l$ case, the completely cooled short loops may absorb the energy from the bottom of the surrounding long loops. 
Looking at the $y-$direction velocity $v_y$ distribution (panel (g)), it is evident that in the region of minimal pressure, both ffHD$_l$ and MHD$_l$ cases fill this pressure minimum by the siphon effect-induced plasma counterflows.
We note in Figure~\ref{fig5}(g) that a region with negative vertical flows is realized through the sharpest density-temperature variations. 
This counterflow region in the MHD$_l$ case is much larger than in the ffHD$_l$ case, meaning it needs to replenish more matter. 
Looking at panel (h), the dotted line describes the direction of the magnetic field ($B_x/B_y$), and the dash-dotted line describes the direction of the velocity ($v_x/v_y$), for the MHD$_l$ case.
Under the ideal MHD assumption of frozen-in plasma, $v_x/v_y$ should strictly coincide with $B_x/B_y$.
However, we observe that at about $y=7.1$~Mm, $v_x/v_y$ deviates significantly from $B_x/B_y$, indicating a strong flow deviating from the magnetic field lines in this region, which coincides precisely with the bottom of the transition region where the temperature starts to rise. 
Therefore, we believe that this is another important reason why the transition region in the MHD model continues to rise during the relaxation process.
Note that, the spikes in Figure \ref{fig5}(h) only located in a very small region.
Therefore, it might be somehow related to the numerical resolution or schemes using here.

In summary, the differences between the two models are all related to the physically allowed mobility of magnetic field lines in the MHD model, which manifests itself already during a relaxation phase and determines the height of the transition region in the models we then start to evaporate towards condensations.
However, the main difference between the two models is not in the lowest chromospheric layers but starts from the transition region, and is especially pronounced at the bottom of the transition region. 
For models with optically thin radiative cooling, the handling of the low temperature limit can never be accurate.
Considering the full radiative transfer may help to improve the results.

Though not related, we have to point out that, the transition region here is higher then values expected from (semi-)empirical models \citep[for example, ][]{vern1981}.
The reason still locates in Figure \ref{fig5}(e).
The heating and cooling in the current model is not perfect balance.
A net energy loss always exists at the top the of transition region.
Such a energy unbalance will cause the transition region to grow higher and higher.
Such a drawback cannot be prevented for a simulation lasting for hundreds of minutes.

\subsection{Full MHD with different field strength}
\label{sec43}
From the above analysis one can conclude that the difference between the ffHD and MHD model depends on the plasma $\beta$, and one may even suspect that if the magnetic field strength $B_0$ in the MHD model were infinitely large, the deviation between the ffHD and MHD model would no longer occur.
That means, the MHD model would gradually converge to the ffHD model as the magnetic field strength $B_0$ increases, since our initial magnetic field before relaxation is a potential field with zero Lorentz force.
We now investigate whether this is really the case.

\begin{figure*}
  \centering
  \includegraphics[width=0.99\linewidth]{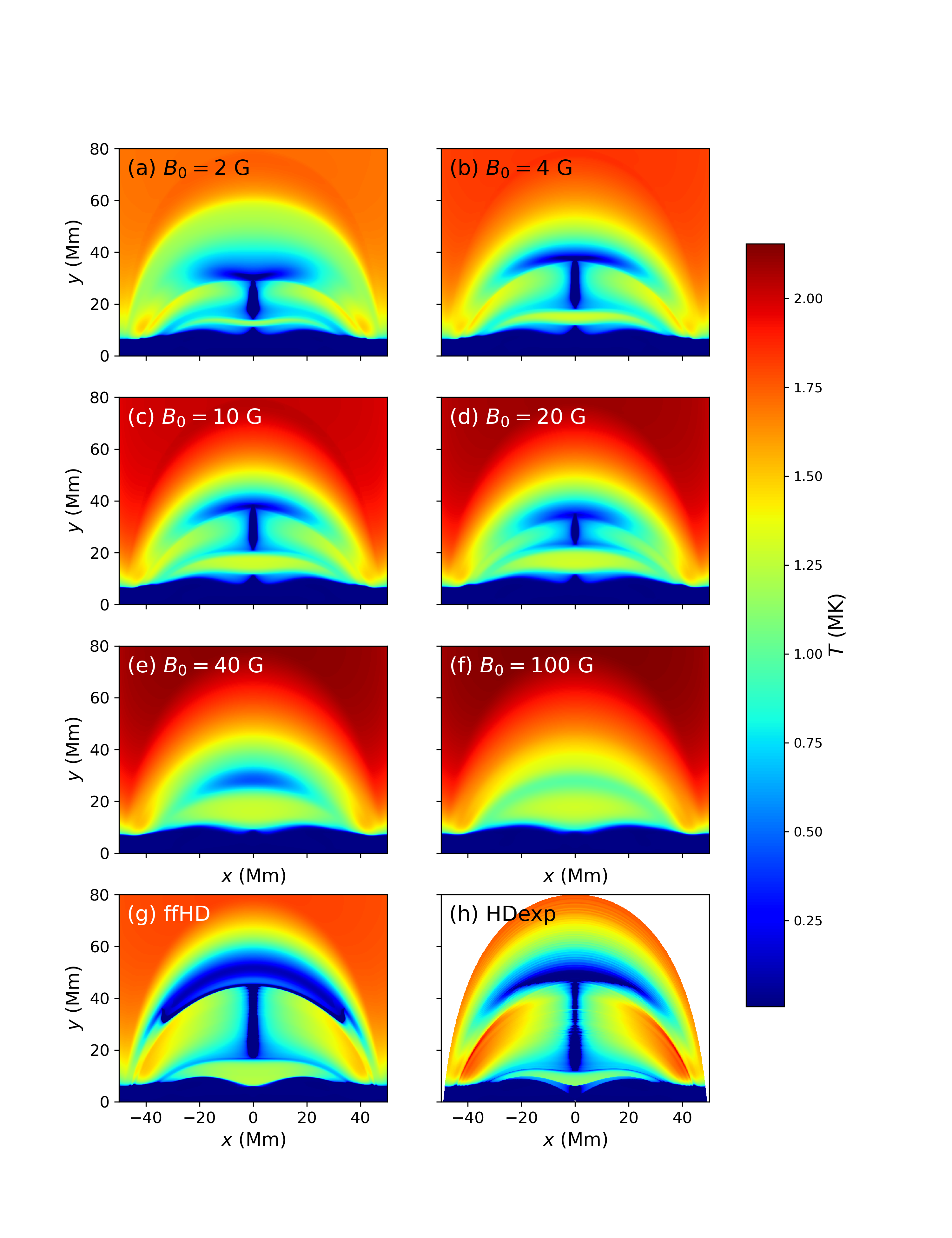}
  \caption{Temperature distribution at $t=143$ min for MHD simulations with $B_0=$(a) 2 G, (b) 4 G, (c) 10 G, (d) 20 G, (e) 40 G, (f) 100 G.
  Panel(a) is the same with Figure~\ref{fig1}(b2).
   {We also add Figure~\ref{fig1}(a2) and (d2) here as panels (g) and (h) for better comparison.}} 
  \label{fig6}
\end{figure*}

In the previously shown MHD simulations, we set $B_0$ to 2 G. 
Now, with the same settings, we conduct five additional sets of MHD simulations with $B_0=4$~G, 10~G, 20~G, 40~G, and 100~G for comparison. 
Figure \ref{fig6}(a)--(f) displays the temperature distributions at $t=143$ min for these six simulations.
 {While panels (g) and (h) are the results from ffHD and HDexp, which are exactly the same with Figure~\ref{fig1}(a2) and (d2).
They are here for an easier comparison with the previous six panels.}
It is noticeable that the stronger the magnetic field, the more difficult it is for condensations to form, or at least the slower the formation of condensation occurs, contrary to our earlier speculation. 
If compared with the previous ffHD model (Figure \ref{fig1}(a2)), it even appears that the weaker the magnetic field is, the closer the MHD result is to the ffHD scenario.
Why is this?

Firstly, we see from Figure~\ref{fig6} that particularly in the outer coronal loops, there appears to be a correlation where stronger magnetic fields correspond to higher temperatures. One may argue that there is a role for more effective numerical dissipation in stronger magnetic fields, resulting in more substantial heating of the corona.
For complicated magnetic field structures with strong and localized magnetic field variations, this could be plausible. 
However, the magnetic field that we initiate here is a fairly smoothly varying potential field, with zero magnetic free energy, already at its lowest energy state.
Therefore, the available energy for dissipation and associated numerical conversion from magnetic to thermal energy is quite limited, and as the end state stays close to force-free, this is rather unlikely.
We propose that a more likely explanation for the observed correlation between stronger magnetic fields and higher coronal temperatures is a slower rate of condensation as obtained near transition region heights for stronger fields.
Because the details through the transition region are intricately influenced by the numerical algorithmic approach taken, our hypothesis suggests that this evaporation outcome is resolution-dependent.

\begin{figure*}
  \centering
  \includegraphics[width=0.99\linewidth]{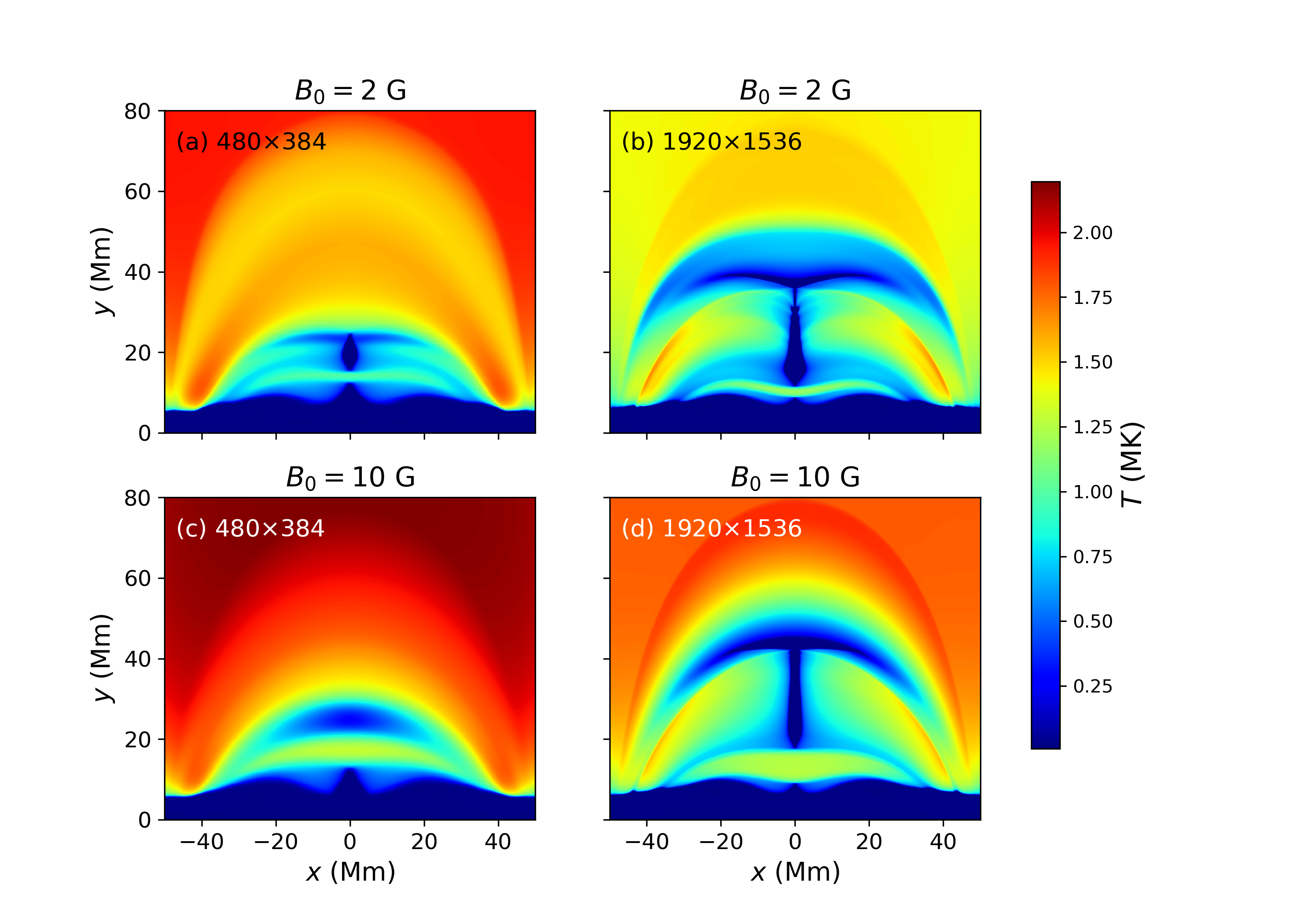}
  \caption{Temperature distribution at $t=143$ min for the MHD model with (a)$B_0=2$~G and low resolution, (b)$B_0=2$~G and high resolution, (c)$B_0=10$~G and low resolution, (d)$B_0=10$~G and high resolution.} 
  \label{fig7}
\end{figure*}

In Figure \ref{fig7}, we present a comparative analysis of two cases from Figure \ref{fig6}, specifically for the $B_0=2$ G and $B_0=10$ G cases, where we have modified the resolution, halving it to 480$\times$384 and doubling it to 1920$\times$1536, and rerun the simulations through both the relaxation and evaporation phases.
We label these new simulations as lor2, hir2, lor10, hir10 cases, respectively.
Panels (a) and (b) show the temperature distribution for lor2 and hir2 cases at $t=143$ min, while panels (c) and (d) display the corresponding temperature distribution for lor10 and hir10 cases.

Comparing with the 960$\times$768 resolution cases in Figure \ref{fig6}(a) and (c), two characteristics are evident. 
First, higher resolutions facilitate earlier condensation formation. 
Second, higher resolutions result in lower coronal temperatures.
These are consistent with previous 1D studies \citep{brad2013}.
However, this also is similar to the characteristics observed with stronger magnetic fields. 

Therefore, we propose that for weaker magnetic fields, a true solution-like state can be approached with relatively low resolutions; in contrast, stronger magnetic fields require higher resolutions to achieve similar fidelity. 
It appears that the process of thermal instability and its occurrence is rather resolution-dependent, with higher resolutions tending towards the true solution. 
In that respect, it is to be noted that early 1D studies \citep[for exmaple, ][]{xia2011, john2019b} used AMR capabilities to facilitate the proper resolution of strong and suddenly forming gradients in all thermodynamic variables.
Here, we did not activate AMR at all, in order to isolate all effects at play in such scenarios, and we did not activate the various TRAC treatments \citep{john2019, zhou2021} invented more recently to handle more correct evaporation fluxes across potentially underresolved transition regions. 

\begin{figure*}
  \centering
  \includegraphics[width=0.99\linewidth]{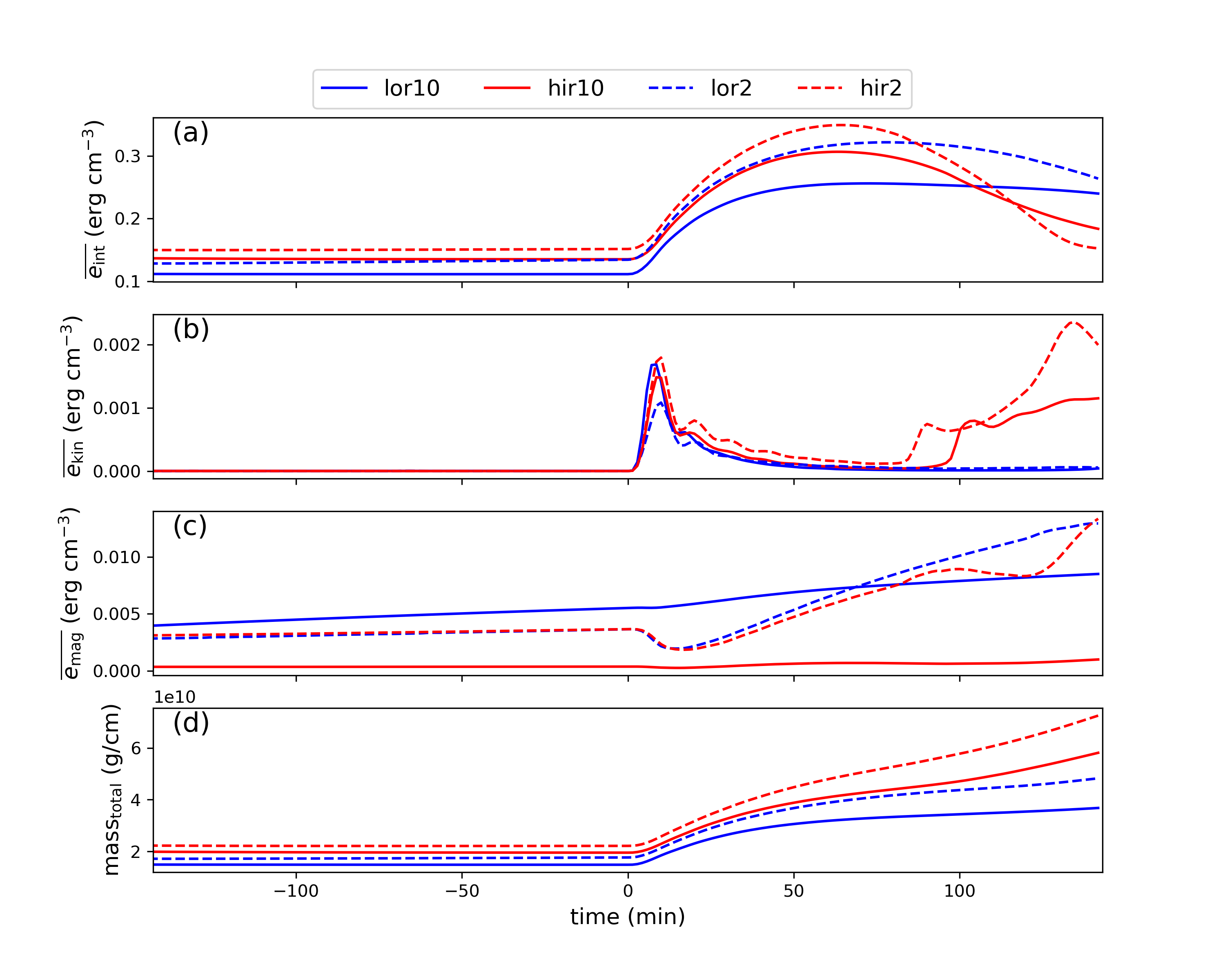}
  \caption{Time evolution of (a) averaged internal energy density, (b) averaged kinetic energy density, (c) averaged magnetic free energy density and (d) total mass in the coronal region for the four cases shown in Figure \ref{fig7}, which represent low (blue) and high (red) resolution cases of $B_0=2$~G (dashed) and $B_0=10$ G (solid) 2D MHD runs.}
  \label{fig8}
\end{figure*}

To further support this claim, Figure \ref{fig8}(a)--(c) displays the temporal variation of the averaged internal energy density, kinetic energy density, and magnetic free energy density in the coronal region ($y > 10$ Mm) for the four simulation sets from Figure \ref{fig7}.
Here, solid lines represent the 10 G cases, dashed lines the 2 G cases, with red indicating high resolution and blue representing low resolution, respectively.
During the relaxation phase, the internal energy evolution is of greatest interest as it far exceeds the kinetic and magnetic free energy.
With no significant difference in coronal temperatures, the lower internal energy that we get at lower resolutions implies a lower coronal density, which in turn is less conducive to evaporation. 
We observe that at the same resolution, stronger magnetic fields correspond to lower internal energy; similarly, for the same magnetic field strength, lower resolutions correlate with lower internal energy, aligning with our earlier explanation.
Although constituting a smaller proportion, the magnetic free energy is also noteworthy, and there is more free magnetic energy available for lower field strength, or for lower resolution in a higher field case.
The magnetic free energy in all cases are exactly zero at the beginning of the simulation.
Then, as mentioned in the previous section, during relaxation, the magnetic field undergoes a slight expansion, gaining some magnetic free energy. 
This is an inevitable result of the interaction between the vertical pressure gradient force and the non-vertically aligned magnetic field. 
However, for the weak magnetic field cases of lor2 and hir2, similar amounts of magnetic free energy are obtained prior to $t=0$, indicating insensitivity to resolution. 
In contrast, for the strong magnetic field cases of lor10 and hir10, the difference is significant. The hir10 case gains almost no magnetic free energy, whereas lor10 even acquires more free magnetic energy than both weak magnetic field scenarios.

After the onset of evaporation at $t=0$, we see in Figure~\ref{fig8} that the density begins to rise, and the total coronal mass change is shown in panel (d).
It is evident that the coronal density during the relaxation phase largely determines the increase in coronal mass during evaporation. 
It can be seen that the low-resolution cases show no significant difference from the high-resolution cases in terms of total coronal mass, particularly in average internal energy, yet the formation of condensation is much slower. 
This again supports our proposed explanation that higher field strength cases requiring more resolution to recover the actual MHD evolution.

Hence, we noticed that the precise time of occurrence of thermal instability is related to the resolution of the simulation, and that a stronger magnetic field requires higher resolution. 
The sudden onset of thermal instability is triggered by small seed perturbations in temperature and density which subsequently grow exponentially as exemplified by linear MHD theory \citep{park1953, clae2020}.
However, these seed perturbations need to be resolved by the simulation. 
Simulations performed at lower resolution more easily smooth out or numerically dissipate these small perturbations, hindering the onset of thermal instability. 
We argue that a similar effect may occur with strong magnetic fields, since the amplitude of evaporation-induced temperature or density perturbations along the magnetic field is inversely proportional to the magnetic field strength.
Adopting the same heating profiles in a stronger magnetic field produces comparatively smaller amplitude perturbations, making it more challenging to propagate these to the corona and cause thermal instability.
Note that the challenge to numerically resolve even tiny fluctuations is aided greatly by AMR, which we typically use in our simulations of multi-dimensional evaporation-condensation driven evolutions \citep{xia2012, fang2013, kepp2014, xia2016, xia2017, zhou2020, li2022}.

\section{Conclusion}
\label{sec5}
Simulating large-scale coronal condensation events by solving the 3D non-adiabatic MHD equations still requires substantial computational resources, even with current technology. 
For certain phenomena like accelerated electron-beam driven evaporation flows in flare loops, or for static loops showing coronal rain events and where one can argue that the dynamics of the magnetic field are not significant, we can feasibly adopt a traditional 1D HD simulation approach. 
In this approach, the magnetic field is approximated as a rigid body, and the plasma flowing along it is treated as a hydrodynamic fluid.
This has the distinct advantage that one can use extreme resolutions easily, and add in effects beyond those we discussed in this paper, such as non-local-thermal equilibrium radiation, or certain non-thermal particle effects.

In this paper, we introduced the ffHD model first proposed by \citet{mok2005} to the open-source MPI-AMRVAC code and conducted several tests. 
We performed a thorough comparison between pseudo-2D and actual 2D ffHD and full MHD simulations, specifically targeted to evaporation-condensation scenarios for filament formation.
The test results indicate that ffHD can indeed replicate most of the outcomes of full MHD simulations. This method can significantly reduce the computational resources required for full MHD simulations.
We showed how the ffHD approach closely follows the pseudo-2D simulations, as long as the latter incorporate the variation of the flux tube cross-section. This geometric effect was also quantified for a large set of 1D runs, by performing both constant area and varying area runs  {on the flux tube} along a representative magnetic field line. We highlighted that more resolution is an absolute necessity for reliably handling evaporation-condensation studies in multi-D MHD setups, especially for higher magnetic field regimes (lower plasma $\beta$), which is significantly aided when the software can perform automated grid refinement.

The paper also provides a detailed comparative study between ffHD and other models, contributing to a better understanding of their differences. 
In our accompanying paper, we will apply the ffHD model to 3D analytical and observational magnetic structures using AMR grids to further investigate the evaporation-condensation mechanism.
\section*{acknowledgments}
       {We thank the referee for valuable suggestions.} YZ acknowledges funding from Research Foundation – Flanders FWO under the project number 1256423N.
      Visualisations used the open source  {Python package  \href{https://matplotlib.org/stable/}{Matplotlib}}.
      Resources and services used in this work were provided by the VSC (Flemish Supercomputer Center), funded by the Research Foundation - Flanders (FWO) and the Flemish Government.
      RK is supported by Internal Funds KU Leuven through the project C14/19/089 TRACESpace, an FWO project G0B4521N, and funding from the European Research Council (ERC) under the European Union Horizon 2020 research and innovation program (grant agreement No. 833251 PROMINENT ERC-ADG 2018). RK acknowledges the International Space Science Institute (ISSI) in Bern, ISSI team project \#545.
\vspace{5mm}




\bibliography{main}{}
\bibliographystyle{aasjournal}


\end{document}